\documentstyle[11pt]{article}
\input epsf
\begin{document}

\begin{center}
{\large{\bf Construction of a natural partition of incomplete horseshoes}}
\\[3cm] \end{center}

\begin{center}
{\bf C. Jung $^{1}$ and A. Emmanouilidou $^{2}$}

$^1$ Centro de Ciencias Fisicas, UNAM, Apdo postal 48-3,
 62251 Cuernavaca, Mexico

$^2$Max Planck Institute for the Physics of Complex Systems,
N$\ddot{o}$thnitzer Stra$\beta$e 38, 01187 Dresden, Germany 
\\[2cm]

{\Large \bf Abstract} \end{center}
A method is presented to construct a partition of an incomplete horseshoe
in a Poincare map which is only based on unstable manifolds of outermost 
fixed points and eventually their limits. Thereby this partition becomes
natural from the point of view of asymptotic scattering observations.
The symbolic dynamics derived from this partition coincides with the one
derived from the hierarchical structure of the singularities of scattering
functions. \\

PACS number(s): 05.45.-a

\newpage
\section{Introduction}
% Ab.1

The most compact global description of the qualitative structure of all
trajectories of a chaotic system is by giving a symbolic dynamics together with
a corresponding overshadowing by basic trajectory segments. To a given
symbolic dynamics there corresponds a partition of that part of the
phase space which covers the chaotic invariant set. Then any
trajectory is characterized by the symbol string corresponding to the
sequence of partition cells visited by this trajectory. The symbolic encoding
of any trajectory as a sequence of symbols allows for computing measures of
chaos using the thermodynamic formalism \cite{Sinai}. In most cases 
it is easier
to work with maps than with flows. Therefore we cast the system under 
study into some iterated map and in this map the chaotic set is represented by
some version of Smale's horseshoe \cite{smale}. Then the construction of
a symbolic dynamics for the system is converted into the task to construct
a partition of that part of the domain of the map which covers the chaotic
set i.e. the fundamental area of the horseshoe. For simplicity in this
article we only consider maps on a two dimensional domain. We think of
Hamiltonian systems with two degrees of freedom which are converted into
a two dimensional Poincare map.

If the horseshoe is complete then the partition is immediate and is
done by a few arcs of the unstable manifolds of some of the fixed points
of the map. The
corresponding symbolic dynamics becomes a complete shift of symbol strings,
where to each partition cell corresponds one symbol value. 
Problems appear if the horseshoe becomes incomplete, if pruning sets in 
\cite{cita} and the symbolic dynamics needs grammatical rules. Then
at first sight there is no natural partition of the horseshoe area which
leads to some useful symbolic dynamics. Some time ago \cite{gras1}-
\cite{gras2} it has
been proposed to use in this case as division lines the line of maximal folding of the
horseshoe. It is essentially the line along which the tip of inner arcs of
unstable manifolds move when we deform a complete horseshoe into the
incomplete one under study by varying some parameter of the system. Later
some criticism of the original method has been published \cite{cri1}-
\cite{cri2} and in reaction to these remarks an improved version of this basic idea has
been
developed \cite{pol1} - \cite{pol2}. This improved version includes
symmetry lines as parts of the division lines and is able to handle KAM
islands in the horseshoe. If the task is only to construct some partition
which separates trajectories, then the problem is solved by this method.

However, any method based on fold lines or, more generally, based on lines
which are not given by invariant manifolds of outermost fixed points of the
horseshoe leads to a completely different type of problems if we use it
for the description of scattering processes \cite{Lipp}. Then the branching
tree coming from the symbolic dynamics does not coincide with the branching
tree coming from the hierarchical structure of the singularities of the
scattering functions. This structure of the scattering functions is 
completely determined by the intersection pattern between the stable manifolds
and the local segment of the unstable manifold of the outermost fixed points
of the Poincare map. Level by level ( iteration step of the map by iteration
step of the map ) new gaps are cut out of this local segment of the unstable
manifold and these gaps correspond one to one to the intervals of continuity
seen in scattering functions. Therefore the partition of the horseshoe becomes
incompatible as soon as it uses any kind of division lines which are not
related directly to these invariant manifolds of the outermost fixed
points. Now the problem arises: Can we construct some partition of incomplete
horseshoes which is exclusively based on this manifold structure and
which is compatible with the branching tree derived from scattering data ?

In \cite{Lipp} an ad hoc solution has been given for a particular case. In 
the present article we show how the basic ideas for this particular solution 
can be converted 
into a general method applicable to any incomplete horseshoe. Of course, any
scattering based treatment of a system can only take into account that part of
the system which is accessible to the outside world and which can be seen
by an asymptotic observer. Therefore our partition only takes into account the
outer unstable part of the horseshoe. It ignores completely KAM islands and
their interior. Near the fractal surface of KAM island and its surrounding
of secondary structures it becomes approximate. We think that an approximate
symbolic dynamics for a chaotic system is as valuable as an approximate
analytic solution for a system which is mainly regular. This is our motivation
for the detailed presentation of the method. 

\section{Horseshoe development stage and its appearance in scattering data}
% Ab.2

In the domain of the Poincare map $P$ the horseshoe is traced out 
by the invariant manifolds of the outer most fixed
points. To start the horseshoe construction 
one defines a fundamental area $R$ which covers the whole chaotic 
invariant set. The choice of $R$ is not unique. We choose $R$ to be
a quadrilateral whose corner points are the outer most fixed point(s)
as well as other primary intersection points \cite{Wiggins} of the invariant 
manifolds. Its boundary lines are segments of the invariant manifolds of the
outer fixed point(s). Under iterated applications of the map   
the area $R$ stretches and folds resulting in a horseshoe.

The image $P(R)$ does not cover $R$ completely. It leaves some holes.
Let us assume that the complement $R \setminus P(R)$ has $n-1$ connected
components. Then we deal with an n-ary horseshoe having n arms.
We call the various components of this complement the first
level unstable gaps $g_{1,j}$ where $j$ runs from $1$ to $n-1$. The unstable 
gaps $g_{k,j}$ of level $k$ are the images of the first level gaps under
the $(k-1)$ fold application of $P$. Stable gaps can be defined in analogy
by using the inverse map. We mainly deal with the unstable gaps, therefore,
if we only say gaps we mean the unstable ones. Higher level gaps can penetrate 
$R$ several times completely and/or several times incompletely.

The boundaries of unstable/stable gaps are given by segments
of the unstable/stable manifolds of the outer fixed point(s). The gaps are areas 
that are not needed to cover the chaotic invariant set. A point that
lies in a stable/unstable gap of level $n$ is mapped out of $R$ after $n$
applications of
the map/inverse map. As we later show, we build the cells for partitioning
the phase space around these unstable gaps.
 
 The development stage of the horseshoe can be
described by parameters, $\alpha_j$, that measure how much the gaps $g_{1,j}$
penetrate through $R$ compared to the complete case \cite{Ruckerl}. For a
complete horseshoe, $g_{1,j}$ penetrate completely through $R$ and $\alpha_j=1$
for all $j$.
 The significance of these parameters is that they describe the qualitative
global
aspects of the outer part of the horseshoe which is accessible to scattering 
trajectories. They ignore the details and in particular they can not describe
the
secondary structures close to the surface of KAM island nor the interior of
KAM islands. However, it determines quite well the outer hyperbolic
part of the chaotic invariant set whose effects are contained in the scattering
data.
To describe the development stage of a horseshoe one needs, in general, a
parameter
$\alpha_j$ for every gap of level $1$.
 If the tip of $g_{1,j}$ ends inside of a stable gap ( and for
arbitrary physical parameters of the system this happens with probability one ),
i.e. if the map avoids primary homoclinic tangencies and if as a consequence
starting 
from a certain level ( call it $l+1$ ) all tips of gaps end outside of $R$, then
each $\alpha_j$ can be expressed as $k_j/n^l$ \cite{Ruckerl}, where $n$ is the
number of arms of the horseshoe defined above ( i.e. if we have a n-ary 
horseshoe ), and $k_j$, $l$ are integers ($k_j$ does not contain any
factor $n$). If the system has symmetries, a smaller 
number of development parameters may be sufficient. For more explanation of this
description of the development stage see \cite{Lipp1}.

Here we give a very brief description how we get the number $k_j$. For
simplicity
we do it for the binary case where only a single gap of level $1$ exists and where
we do not need the index $j$. Let us assume that the first level unstable gap
ends
in some segment of the l-th level stable gap. Then this segment of the gap
corresponds to
some segment of the l-th level stable gap of the corresponding complete
horseshoe.
Then we take in the complete horseshoe all gaps up to level $l$ and count all
their
penetrating segments ( in the complete case there are no partial penetrations ) 
consecutively starting from the side of the outer fixed point. The number $k$ is
the number, in the complete case, of that segment in which the unstable first
level gap ends, in the
incomplete case under consideration ( for more details see \cite{Lipp1}). 
Note that not all numbers of the form $k/n^l$ correspond to existing development
stages of horseshoes. In \cite{Ruckerl} it has been explained why by discussing,
 as an illustrative example, the nonexistence of the binary $3/8$ case.

Our method for partitioning the phase space requires the knowledge of these
development parameter(s). 
 If only asymptotic data are available, then, $\alpha_j$ can be 
obtained from the symbolic encoding of the hierarchical structure
of the fractal set of singularities of the scattering function
\cite{Lipp1}-\cite{tj}. 
Sometimes symmetry considerations must be included, as well, to decide the basic 
type of the horseshoe, e.g. whether it is a binary or a ternary one.  

\section{Algorithm for partitioning the phase space}
% Ab.3

To avoid clumsy notation we present our method for the case of only one gap of
level $1$.
 If the map has several ones, then for each one of them the same procedure has
to be 
performed. In the examples we will also show ternary horseshoes with 2 gaps of
level $1$.
First we make the following definitions for the notation:
 From now on we call such segments of gaps of level $p$ which cut $R$ completely 
the $A_p$ part and ( if it exists ) the remaining part which only penetrates
partially into $R$ the $B_p$ part ( note that $B_p$ is always the image of a
part
of $B_{p-1}$. Then there exists an integer $m$ such that for levels 
$1 \le p \le m$ the whole gap $g_p$ only consists of one $B_p$ part and the
level 
$m+1$ is the first level for which the gap contains an $A$ part and cuts $R$
completely.

 At every level $p$ we define the unresolved region $U_{p}$
as that part of the fundamental area $R$ which is not yet
assigned to any partition cell. In general, $U_{p}$, might consist
of various components.

Before we give rules in the form of an algorithm, for partitioning the phase space, let us emphasize two principles
which
we must follow strictly in order to obtain a partition which coincides with the
one 
seen by the asymptotic observer: \\
{\bf Principle 1 : } If an arc of the unstable manifold ( i.e. boundary of a gap
)
leaves $R$ and reenters $R$ then the two cells which this arc connects must be 
different, i.e. must belong to different symbol values, always if the preimage
of 
the segment which leaves $R$ has completely been inside of $R$. The arc of the
unstable manifold under consideration may connect the boundary of two different
cells. \\
{\bf Principle 2 : } There must be a division line between two areas inside of
$R$ 
where the
arcs of unstable gaps have qualitatively different behaviour. Thereby we mean:
$R$ has two stable sides and there are segments of unstable gaps in $R$ which
connect
one of these stable sides with the other one. There are also segments of
unstable gaps
which start on one of these sides and return to the same side. Such arcs wind
around
some incomplete unstable cut ( $B$ part of a gap ) of lower level. The boundary
between 
areas of these two types of behaviour must become a division line. In addition
if some arcs wind around a $B_p$ gap and other arcs wind around a different
$B_q$ gap then these two areas also must be separated by a division line. \\
In total we obtain three types of division lines. First, images of $B$ gaps (
division
lines of class $1$ ).
Second, boundaries of areas where arcs wind around a particular $B$ gap.
The first class of division lines is defined on finite level of the hierarchy,
the second case is in many cases defined by an iterative procedure in
the limit of level to infinity only ( such division lines are of class $2$ ). 
Sometimes ( it allways happens in hyperbolic incomplete cases, but not only in
them ) 
this division is done at a finite level ( division
lines of class $3$ ). This happens if arcs of gaps make turns such that the outer
boundary leaves $R$ whereas the inner boundary turns inside of $R$. 
If division lines of class $3$ do not occur, then the partition
leads to grammatical rules of length $1$, if division lines of class $3$ exist, then
the grammar may be of length larger than $1$. To get a grammar of length $1$ in such
cases
may require the introduction of additional division lines. They are $A$ parts of 
preimages of 
division lines of class $3$ and thereby are gaps in whose interior some tip of a
stable
gap ends. See the examples in the next section. Division lines of class $1$ can
appear 
up to level $l+1$, division lines of class $3$ can appear up to level $2l$.\\
The algorithm can be cast into the following steps:\\
{\bf 1.} Start with the fundamental rectangle
 $R$. It has $2$ unstable sides, at least one of them is a local segment of an
unstable
manifold of an outer fixed point. The $2$ unstable sides of $R$ will later become
the outer boundaries of $2$ partition cells. Already now we can assign to these
boundaries the symbol values which will be assigned later to the adjacent cells.
For boundaries which are local segments of unstable manifolds we use as symbol
values
the names of the corresponding outer fixed points, such that the corresponding
cells
represent these fixed points in the sense of overshadowing. \\
{\bf 2.} Define the integer $m$ as the one fulfilling the relation
\begin{equation} 
\frac{1}{n^{m}}\le\alpha< \frac{1}{n^{m-1}}
\end{equation}
% Gl.1
for $\alpha < 1$. In the particular case of $\alpha = 1$ we set $m=0$.
It coincides with the $m$ mentioned above.
Extend the unstable manifold level by level, i.e. construct the gaps
$g_p$ for $p$ from $1$ up to $m$. None of these gaps will penetrate $R$ completely,
i.e. none will cut $R$ into several components.
 Each $B$ part of a gap $g_p$ will become the center of
a partition cell ( thereby principle 2 will be fulfilled ). 
We already assign the corresponding symbol value to
these gaps during the construction of the levels from $1$ to $m$. Accordingly at
level $m$ we have already $m+2$ different symbol values in total ( including
the two symbols for the unstable sides of $R$ ). \\
{\bf 3.} The unstable gap $g_{m+1}$ is the first gap that cuts through $R$
completely
and separates $R$ into at least two disconnected parts. All $A_{m+1}$ cuts must
become 
the division lines between different cells. At least one of the parts into which
$R$ is 
cut is a topological rectangle. At the moment we take it as one complete cell.
Eventually it may be cut into several ones at higher levels. 
 At the inner side of the cuts we assign a new symbol value which belongs
to a partition cell which will be constructed on this side on higher levels. 
This step is a special case of the more general step 5.C explained below. \\
{\bf 4.} If $k \ne 1$ then $g_{m+1}$ contains a $B_{m+1}$ part in addition.
As in step 2 this $B$ part becomes the center of a new partition cell and we
have to give it a new symbol value. \\
{\bf 5.} Starting from level $m+1$ each gap cuts through $R$ completely,
in general several times. And up to level $l$ it contains in addition one $B_p$
part
close to its tip. These $B_p$ parts we treat as before: We assign a new symbol
value 
to the new unstable boundary around $B_p$. It will become the outer unstable
boundary of
a partition cell. The inner boundary of the corresponding cell will be
constructed at 
higher levels. For the various $A_p$ cuts we must distinguish several
possibilities: \\
{\bf A.} Such cutting segments are images of $A_{p-1}$ and run in the interior
of
already existing cells. Then they do not have any function for the partition at
this moment.
However they may be turned into division lines later, if point 7 applies to
them.\\
{\bf B.} Such cutting segments are images of $A_{p-1}$ and run through
$U_{p-1}$.
They cut $U_{p-1}$ into two components and one of them ( we call it the outer
component )
is a quadrilateral.
Then, the same symbol value holds on both sides of the cutting segment. The cell
that already exists on the outer side of this cutting segment is extended
towards
the inner side and its inner boundary will be improved on higher levels. 
 However, it must become a division 
line if it falls into the class $3$ mentioned in the discussion of principle 2. 
If all parts of $U_p$ should be topological rectangles then they become partition
cells
and the process stops. This happens for hyperbolic cases and for hyperbolic 
incomplete horseshoes this final step and its corresponding decomposition of
 $U$ into quadrilaterals necessarily includes division lines of class $3$. \\
{\bf C.} Such cutting segments $A_p$ are the image of the $B_{p-1}$ gap.
Then this $A_p$ segment becomes the division line between two different cells in
any case. Now we must distinguish three subpossibilities: \\
 --a) If the cutting segment cuts off from $U_{p-1}$ areas that are 
topological rectangles, then
 these rectangles are complete cells and are labelled by  
the symbol value which already existed on the opposite unstable side of these 
rectangles. 
We introduce a new symbol value only on the side of the
cutting segment which is not a topological rectangle
and start constructing a new cell whose outer boundary
is the cutting segment under consideration.\\ 
 --b) If the cutting segment does not cut off from $U_{p-1}$ a
topological rectangle ( i.e. if on both sides polygons remain with more
than four corners ) 
 then, assign a new symbol on each side of the cutting segment under
consideration. The inner boundaries of the corresponding cells
will be constructed at higher levels. \\
 --c) Also if an image of a $B$ part cuts an already existing cell into several
pieces, then these pieces must become independent cells and must have different
symbol values. \\
{\bf 6.} At levels $p>l+1$, the cutting segments of the $p$ level gap are always 
images of $A_{p-1}$ gaps. Thus, we proceed to higher levels by iteratively
applying 
step 5.B to improve the cell boundaries iteratively. \\ 
{\bf 7. } If division lines of class $3$ exist and if we are interested in
grammatical rules of length 1 then finally we convert gaps of type $A$ into
division lines, if they fulfil the following properties: They contain tips of 
stable gaps. This is equivalent to being a preimage of a division line of class
3
such that the part of the division line of class $3$ which leaves $R$ has a
preimage
completely inside of $R$. There can only be a finite number of such additional
division lines.
Note that points 3 and 4 are special cases of point 5. We took them as separate
points
to start with simpler special cases. If we only want a minimal set of
instructions
for the division then we can drop points 3 and 4.

 Note that step 5 is basically the realization of our guiding principle 2 and
therefore it assures that scattering trajectories belonging to different
intervals
of continuity of the scattering functions are encoded
by different symbol sequences. Case 5.B. accounts for trajectories that have
been
separated at a previous level while step 5.C. accounts for those that will be
 separated at the current level. 

The grammatical rules of length one corresponding to this division of $R$ into
cells
are obtained by observing  how any cell is covered by the
images of the other cells.

\section{Examples}
% Ab.4

The procedure will become clearer by doing a few examples in detail.
 In the following plots
thick solid lines are final division lines of cells defined on a finite level.
 Thick broken lines are division lines
in the limit of level to infinity but without taking into account nonhyperbolic
effects. Dot dashed thick lines are division lines according to step 7 and
 division lines of class $3$.

\subsection{ Binary case $\alpha = 1/2$ }
% Ab 4A

 We illustrate our algorithm first for a binary horseshoe with $\alpha=1/2$. 
In this case, $m=1$, $k=1$ and $l=1$. For a binary
 horseshoe, the map has two fixed points, one of them being the outer fixed
point. 
As a Poincare map we use the model map also used in \cite{Ruckerl}:
\[
x(n+1) =x(n)+p(n), 
\]
\begin{equation}
p(n+1) =p(n)+A f(x(n+1))
\end{equation}
% Gl.2
Here $x$ is the position coordinate, $p$ is the momentum coordinate,
 $n$ is the discrete time, and $A$ is a free parameter. We take the
force function to be: 
\begin{equation}
f(x)=x(x-1)e^{-x}.
\end{equation}
% Gl.3
$A$ is a parameter which can be used to adjust the development stage of the
horseshoe. First we use $A=3.4$ which is close to the lower end of the interval
of $A$ values where a development parameter $\alpha = 1/2$ is realized. As shown 
in \cite{Ruckerl} for this value of $A$ nonhyperbolic effects only start at
very high levels of the hierarchy. Later we change $A$ to get a case of
$\alpha=1/2$ where nonhyperbolic effects set in at rather low levels. In this
later case we briefly discuss what kind of errors our method makes, if nonhyperbolic effects occur.

In Figs.(1a) and (1b) we draw $x$ and $p$ for half integer multiples
of the time step $n$. We choose this Poincare surface of section for
simplicity since with this choice of surface 
the $p\rightarrow -p$ transformation interchanges the stable and unstable manifolds.   
 In Fig.(1a), we draw the stable manifold $W^{s}$ up to level $0$ and the
unstable
one up to level $2$. The topological rectangle $OGHI$ is the area $R$, with
$O$ being the outer fixed point. The gap of level $1$, $g_{1}$, does not cut
through $R$
completely. The gap of level $2$, $g_{2}$, cuts through $R$ completely once. 
Thus, $g_{2}$, cuts $R$ in two disconnected parts.
 One part is adjacent to the outer fixed point, $O$, and is topologically
a rectangle. This part is already a complete partition cell and we assign to
it the symbol $O$, it is the cross hatched area in Fig.(1a). The remaining part
of area 
$R$ is $U_{2}$. $U_{2}$ has three
separate unstable boundary parts. We assign to each one of them a symbol, \%,
 \& and $+$, respectively, which
at higher levels becomes the symbol of the adjacent partition cells. The inner
boundaries of these partition cells are not defined at level $2$. Thereby the
first 5 steps of the algorithm are already done. Note that because the 
$g_2$ gap
ends outside of $R$ there will never be any B parts of gaps starting from level
$2$.
Because $l=m$ in this particular case ( the reason is that $k=1$ ) step 5
is trivial and the rest is an iterative improvement of the inner boundaries
of the cells \%, \& and $+$. From the application of the map we read off the following
grammatical rules: \newline
{\bf 1 :} $+ \rightarrow O$, \%, {\bf2 :}$ O \rightarrow O$, \% , {\bf 3 :} 
\% $\rightarrow $\& , {\bf 4:} \& $\rightarrow +, O$.
  
In Fig.(1b) we draw the boundaries reached at level 6. The binary case 
$\alpha = 1/2$ has a large KAM island around the inner fixed point at $(x=0,p=0)$. The figure
demonstrates how the part of $U_2$ outside of this KAM island is divided into
$3$ cells. 
 In Fig.(1b) we see how iterative division lines separate segments of
different behaviour. Cell \% corresponds to segments which connect the lower boundary
(IH) of $R$
with the upper boundary (OG) and end to the left of $B_1$. Cell $+$ corresponds to
segments which connect the lower boundary of $R$ with the upper one and end to 
the right of $B_1$. Cell \& corresponds to arcs which wind around $B_1$ and
which start and end at the upper boundary of $R$. The behaviour in cell $O$ is
qualitatively the same as in cell $+$ but must be divided because of principle $1$.

Next let us put the parameter $A$ of the interaction in Eq.2 to the
value $4$,
which still is the binary $\alpha = 1/2$ case, but it is close to the upper 
limit of the A interval which belongs to $\alpha=1/2$ and therefore
nonhyperbolic 
effects become important at rather low levels
( see section 4 in \cite{Ruckerl} ).
 Fig.(2a) shows that during the change of the parameter from $A=3.4$ to  $A=4$
a homoclinic bifurcation has appeared. One intersection point of the 
case $A=3.4$ has split into
the 3 intersection points indicated by the little circles in Fig.(2a).
 Since this homoclinic bifurcation shows up 
in an intersection between two tendrils of level $4$ in the interior
of $R$ it results in secondary intersections
of the unstable manifold at level $8$ with the local segment of the stable
manifold ( if any homoclinic intersection point suffers some
bifurcation, then all its images and preimages show the same type of
bifurcation).
This is demonstrated in Fig.(2b), where
the unstable manifold at level $8$ intersects the local segment of the stable
manifold at three points (indicated by squares) giving rise to a secondary
tendril. It acts like a $B$ part of a primary gap and creates a new segment of
unstable boundary of the cell. To take it into account properly in the
symbolic dynamics and in the partition it would be necessary to introduce a
new
cell around this secondary tendril. However, we do not take into account 
these effects, we ignore them and thereby make an error, 
here our construction is approximate. For the creation of secondary tendrils
see
also figures 6 and 7 in \cite{Ruckerl}.

In higher levels of the hierarchy such homoclinic bifurcations become more
frequent
and in the limit of level to infinity they take over in all nonhyperbolic
cases.
Note that they first appear in the vicinity of the large KAM island.
Considering that the divisions between
the cells \&, \%, + end in the surface of the KAM island and dive into its
fractal surroundings makes it evident that the correct partition lines
must approach fractal curves and start to develop wild oscillations if we go
to
high levels of the hierarchy beyond the level where nonhyperbolicities set in.
The oscillations create all the secondary tendrils.
The beginning of such wild oscillations in division lines has been shown in
Fig.8 of Ref. \cite{Lipp} for the ternary symmetric case $\alpha_1 = \alpha_2 =
1/3$.
Our approximation replaces the complicated curve by a smooth curve making
shortcuts
through all fine oscillations.
In all the other nonhyperbolic cases similar effects occur and in the examples 
that follow we do not mention them explicitely.

From this binary $\alpha=1/2$ example it should be clear how all binary $\alpha = 2^{-j}$ cases
are treated. For all these cases, the plot showing how we partition the phase
space at level $l+1 = m+1$ looks similar to Fig.(1a) with the
only difference that $U_{l+1}$ contains $l$ partial penetrations of the gaps
$g_1$ up to $g_{l}$. Accordingly $U_{l+1}$ contains $l+2$ unstable boundary
components in total and we need $l+3$ partition cells in total for the
horseshoe. This observation will become important later in section 5.

\subsection{Binary complete case $\alpha=1$}
% Ab 4B

Of course for this case the gap $g_1$ is the ideal division line of $R$ into two
cells
which are topological rectangles
and it gives the standard symbolic dynamics in 2 symbol values. On the other
hand
we could also take the complete case as the
special $m=0$ case of the sequence of cases mentioned at the end of the previous
subsection. Then $m=0$ and we construct a symbolic dynamics in 3 symbol
values, see Fig.(\ref{fig:binary1}). 
 On level 1 the
gap $g_1$ is the first one which divides $R$ and according to rule 3 we give
two different symbol values on both sides. Formally we can consider this
division line
the image of $B_0$. We see that the side which contains
the outer fixed point is a topological rectangle and assign to it the symbol O.
But we do not recognize that the other half is also a rectangle and assign the
symbol
\% on this side of $g_1$. As third symbol we assign \& to the opposite side 
of
$R$.
Then iteratively the cells \% and \& are enlarged to the inside and in the
limit
of level to infinity their inner boundaries converge from both sides to the
local branches of the unstable 
manifold of the inner fixed point. The grammatical rules are: \newline
{\bf 1 :} $ O \rightarrow O, $\& , {\bf 2 :} $ \% \rightarrow O,$ \&, {\bf 3 :} 
$ \& \rightarrow O, \% $. 

The topological entropy of this symbolic dynamics is
$ln(2)$ and it is a complete binary symbolic dynamics in disguise.
All partition lines consist of vertical lines only which run from one stable
boundary of $R$ to its opposite stable boundary. Of course we could construct 
an equivalent partition by using the corresponding horizontal strips and
division
lines. This shows that in this case global continuous stable and unstable
directions exist and this indicates that this case is a uniform hyperbolic one.

\subsection{Binary $\alpha=7/8$ case}
We now illustrate our algorithm for the binary
$\alpha=7/8$ case using a schematic plot. This is a nice example where
division lines of type $3$ exist. In this case, $m=1$, $k=7$ and $l=3$. In Fig.(\ref{fig:binary78}), we draw 
the stable manifold $W^{s}$ up to level $0$ and the unstable one up to level
$6$. The topological rectangle OGHI is the area $R$, with $O$ being the outer
fixed point. The gap of level $1$, $g_{1}$ does not cut through $R$
completely. The gap of level $2$, $g_{2}$, cuts through $R$ completely once
($A_{2}$ part) and also cuts partially through $R$ ($B_{2}$ part). $A_{2}$
cuts $R$ in two disconnected parts. One part is adjacent to the outer
fixed point, $O$, and is topologically a rectangle. This part is already a
complete partition cell and we assign to it the symbol $O$, it is the cross
hatched area in Fig.({\ref{fig:binary78}). At level $2$, the remaining area has three separate
unstable parts. We assign to each one of them a symbol, $+$, \% and $*$,
respectively, which at higher levels becomes the symbol of the adjacent
partition cells. The inner boundaries of these cells are not defined at level
$2$. At level $3$, the gap $g_{3}$, cuts $R$ completely three times, parts
$A_{3,1}$, $A_{3,2}$ and $A_{3,3}$, respectively, and also cuts partially
through $R$, $B_{3}$ part. $A_{3,1}$ is the image of $A_{2}$ and runs through
an already existing cell, $O$, it is therefore irrelevant. $A_{3,2}$ is
the image of $A_{2}$ and cuts $R$ in two components. One of them is a
quadrilateral and it is already labeled by the symbol $+$. Before though
we assign a symbol value on the other side of $A_{3,2}$ we note that there
are division lines of type $3$ in this case. That is, at level $4$ the gap
$g_{4}$ cuts through $R$ completely $7$ times, parts $A_{4,1}$ up to
$A_{4,7}$. Note that the inner boundary of $A_{4,3}$ and $A_{4,4}$ does not
cut through $R$ while the outer boundary does. Thus, these lines are division
lines of type $3$. In addition, according to step $7$ of our algorithm
the preimage of parts $A_{4,3}$ and $A_{4,4}$, part $A_{3,2}$, is also
a division line. So, to the outer side of $A_{3,2}$ we introduce 
a new symbol value, $4$. Next, $A_{3,3}$ is the image of $B_{2}$ so
it is a division line and we introduce two different symbol values on either
side, \# and $1$, respectively. Next, $B_{3}$ is a new unstable boundary and
we thus introduce a new symbol value, $2$. At level $4$, $A_{4,1}$ and
$A_{4,2}$ run through already existing cells and are thus irrelevant. $A_{4,3}$
 and $A_{4,4}$ are division lines of type $3$, as we have already noted, and
different symbol values must be introduced on either side of them. $A_{4,3}$
cuts on its left side a quadrilateral, thus this side is now a complete
cell already labeled by the symbol $1$. On the other side the symbol \%
already exists. On the left side of $A_{4,4}$ the partition cell \% already
exists and so we just introduce a new symbol value, $6$, on the other side.
 $A_{4,5}$ and $A_{4,6}$ are images of $A_{3,3}$. $A_{4,5}$ cuts on one
side a quadrilateral and so the same symbol holds on both sides,
$*$. $A_{4,6}$ also cuts a quadrilateral and so the same symbol value \#
holds on both sides. $A_{4,7}$ is the image of $B_{3}$ and cuts a
quadrilateral on one side which is already labeled by $4$. We introduce
a new symbol value on the other side of $A_{4,7}$, $5$. At level $l+1=4$
we are through assigning symbol values. But since we have divisions
lines of type $3$ the cells centered around $B_{2}$ and $B_{3}$ take their 
final form only by considering levels up to $2l=6$. In
Fig.(\ref{fig:binary78}), for simplicity, we only indicate those parts of gaps $g_{5}$
and $g_{6}$ that wind around $B_{2}$ and $B_{3}$. These parts are also
division lines of type $3$ and they finalize the cells centered around
$B_{2}$ and $B_{3}$. The grammatical rules are: \\
{\bf 1}: \% $\rightarrow$ *,\&,O, {\bf 2}: * $\rightarrow O$,4, {\bf 3}: \&
$\rightarrow$ \#,2,1, {\bf 4}: O $\rightarrow$
O,+, {\bf 5}: 4 $\rightarrow$ 1, 6, {\bf 6}: \# $\rightarrow$ *, \#, {\bf 7}: 2 $\rightarrow$ 5,4, {\bf 8}: 1 $\rightarrow$
\#, *, {\bf 9}: + $\rightarrow$ \%, {\bf 10}: 5 $\rightarrow$ 1, 6, {\bf 11}: 6
$\rightarrow$ 5,O.

\subsection{Ternary asymmetric case $\alpha_1 = 1/3$, $\alpha_2 = 1$}
% Ab 4D

 In Figs.(5a) and (5b) we draw the invariant manifolds of 
a system that
describes the classical scattering of an electron from a one-dimensional
inverted Gaussian potential in the presence of a strong laser field.
 This potential has offered considerable insight into the laser atom
interactions \cite{Gavrila}. The Hamiltonian in the Kramers-Henneberger
reference 
frame (the frame
which oscillates with a free electron in the time-periodic field) \cite{Kramers}
is:
\begin{equation}
\label{eq:Hamiltonian}
H=\frac{p^2}{2}-V_{0}e^{-((x+\beta_{0}\sin(\phi))/\delta)^2}+\omega I.
\end{equation} 

 Note, that in Eqs.(\ref{eq:Hamiltonian}) we have
transformed to a two-dimensional time-independent system, where
the total energy is conserved. $I$ and $\phi$ are respectively the
action-angle variables of the driving field and $\phi=\omega
t$. $\beta(t)=\beta_{0}\sin(\omega t)=-q E_{0}/\omega^{2}$ is the classical
displacement of a free
electron from its center of oscillation in the time-periodic electric field
 $E(t)=E_{0}\sin(\omega t)$, where $T=2\pi/\omega$ is the period of the
field. The parameters chosen here are $V_{0}=0.27035$, $\delta=2$,
$\omega=0.65$ and $\beta_{0}=0.9$, all given in atomic units (a.u.). The
$\alpha's$ for these
parameter values are $(1/3,1)$. The outer most fixed points
are located at $x\rightarrow \pm \infty$. For the left outer fixed point,
 $A$, we have, $\alpha=1$ and $l=0$, $m=0$. For the right outer fixed point,
 $C$, we have, $\alpha=1/3$ and $l=1$, $m=1$. 
For this driven system, the P-map is a stroboscopic plot. That is, we plot 
$x$ and $p$ every complete period of the field solving Hamilton's equations of
motion. We choose our Poincare surface of section to be $\phi=\pi/2$
because of symmetry reasons \cite{Emmanouilidou}. In Fig.(5a) we draw the stable
manifolds $W^{s}$ up to level $0$ for both outer fixed points, the
unstable $W^{u}$ of the fixed point $A$ up to level $1$, and the unstable
$W^{u}$ of the fixed point $C$ up to level $2$. The 
4-sided area $AECD$ is the area $R$. At this level, the cells adjacent to the
fixed points are already complete ones and we label them $A$ and $C$
respectively. The remaining part of $R$ is the unresolved region and has three
separate unstable boundary parts.
 We assign
to each one of them a symbol, $+$, \& and \% respectively, which at higher
levels becomes the symbol of the adjacent partition cells. The inner boundaries
are not yet defined. In Fig.(5b) we draw the unstable manifolds up to level
$4$ for both fixed points to
demonstrate how the inner boundaries of the cells $+$, \& and \% are defined
at higher levels through an iterative process. In Fig.(5b), we see how
iterative division lines separate segments of different behaviour. Cell $+$
corresponds to segments which connect the lower boundary (AD) of $R$ with the
upper boundary (EC) of $R$ and end to the left of $B_{1}$. Cell \&
corresponds to segments which connect the lower boundary of $R$ with the
upper boundary of $R$ and end to the right of $B_{1}$. Cell \% corresponds to
arcs which wind around $B_{1}$ and which start and end at the lower boundary
of $R$. The behaviour in cell $C$ is qualitatively the same as in cell \& but
must be divided because of principle $1$. The same is true for cells $A$ and
$+$. At level $4$
one strip to the partition cell $+$, two strips to the cell \&, and
three strips to the cell \% have been added. The inner boundaries of these cells
will be 
improved iteratively at still higher levels. From Figs.(5a) and (5b), we find
the grammatical rules to be {\bf 1 :} $A \rightarrow A$, \&, $C$ {\bf 2 :} 
\& $\rightarrow C$, \% {\bf 3 :} $C \rightarrow C$, \% {\bf 4 :} \%
$\rightarrow +, A$ {\bf 5 :} $+ \rightarrow A$, \&, $C$.   
  
In Fig.(5b) we indicate the KAM island around the inner elliptic fixed point $B$
by
an invariant KAM torus quite close to the surface of the island. In the figure
it is
the dotted line. The fixed point $B$ itself at $x=-0.29$ and $p=0$, is marked by
a square.
The KAM islands and their fractal surrounding of secondary structures are the
parts
of the horseshoe which are not treated by our scattering oriented partition.
The existence of KAM islands leads to
 non-hyperbolic effects such as secondary tendrils and related homoclinic
tangencies,
i.e. non transversal intersections, between stable and unstable manifolds
\cite{Ruckerl}. 
 For the parameters we currently consider the
tangencies may show up at level $8$ and higher. Our method does not account for
these additional
intersections. This means that our partition can
not encode the path of those trajectories that ``dive'' into the secondary
structures around 
the fractal surface of stability islands
and stay there for a very long time. However, it successfully accounts
for short and intermediate time scales which are the ones most relevant to 
scattering.

\section{Connection to rotation numbers of the central KAM island 
and to the selfpulsing effect}
% Ab.5
In \cite{jms} it is shown how the selfpulsing effect of chaotic scattering
systems with horseshoes of small development parameters ( where a large 
scale KAM island around a central elliptic fixed point exists ) can be used
for the
inverse scattering problem. The method is based on a simple relation between
the
development parameter and the delay between adjacent pulses. This delay is
just
 the number of steps
of the map needed for a general trajectory in the vicinity of the large KAM
island to encircle this island. In this section we explain how such a
connection
also drops out of our construction of a partition.

For simplicity let us first consider the binary case and start with the
development parameter $\alpha=1/2$ for which we see the figures in section 4.1.
As Fig.(1b) indicates the large KAM island ( containing the inner fixed point 
which is elliptic for $\alpha=1/2$ ) is surrounded by 3 different
partition cells. Points in cell + and close to the boundary of the KAM island
are mapped into cell \% and again close to the boundary of the KAM island.
Such points in turn are mapped into cell \& and again close to the boundary 
of the KAM island. The images of such points lie in cell + and are again
close to the boundary of the KAM island. In total, points close to the 
KAM island need 3 steps of the map to circle once around the KAM island.

Next imagine the binary case $\alpha= 2^{-l}$. According to the remarks at the
end of subsection 4.1 the region $U$ is surrounded by $l$ incomplete gaps
$g_j=B_l$,
$ j=1,...,l$ all being the centers of corresponding partition cells. In
addition there is a cell adjacent to the left hand unstable boundary of $R$
and
one cell adjacent from the left hand side to the gap $g_{l+1}=A_{l+1}$. All
such cells
touch the large KAM island ( in contrast the cell O adjacent to the
outer fixed point is the only cell not neighbouring this KAM island ), i.e.
there are $l+2$ neighbouring cells of the central KAM island. Since $B_p$ is
mapped into $B_{p+1}$ under the Poincare map, it should be clear
that points close to the KAM island are mapped from any one of these cells
into the neighbouring cell in the clockwise orientation. Therefore it takes
$l+2$ steps of the map for such a trajectory to circle once around the
KAM island. 

For all these cases the connection between the number $N_s$ of steps it takes
to 
encircle the KAM island and the development parameter $\alpha$ is
\begin{equation}
N_s = - \log_2 (\alpha) + 2
\end{equation}
Finally, in the spirit of a smooth interpolation we use this equation as
estimate
for the rotation number of the surrounding of the central KAM island for all
small values of $\alpha$. Note that such a formula does not make any sense for
large values of $\alpha$ where no large central KAM island exists. However, we 
could apply an analogous consideration to estimate the rotation number of any
other KAM island appearing for any value of $\alpha$.

In \cite{jms} we have given an estimate of this rotation number which resulted 
in an equation similar to Eq.5 with the only small difference that the
constant
was $3/2$ instead of $2$. Why this small discrepancy ? The explanation is rather
simple: In \cite{jms} we were interested in trajectories in the secondary
structures directly above the surface of the KAM island. In the present
article
we are concerned with trajectories in the globally unstable ( approximately
hyperbolic )
part of the system which ends further outside of the surface of the KAM 
island. In general the rotation number in the KAM island and its secondary 
surrounding is somewhat smaller than in the region further out, therefore the
small
difference in the two estimates.

Let us give a very brief description of the selfpulsing effect itself and some
relation to our partition. Imagine we put some blob of initial conditions into
a single partition cell and close to the central KAM island but outside of it. 
When does a part of the 
trajectories leave $R$? Only the images of the cells around $B_l$ and
$A_{l+1}$
have parts outside of $R$. The other $l$ cells around the island do not have
this property. Therefore only after each complete revolution around the
island, i.e.
 after $N_s$ steps of the map a pulse leaves the inner region and goes out to
the
asymptotic region where in a scattering experiment the time delay $N_s$
between
adjacent pulses can be measured. Then by inverting Eq.5 the asymptotic
observer
can reconstruct the development parameter $\alpha$. For all the details of 
this idea see \cite{jms}.

In analogy to the derivation of Eq.5, for the ternary case the rotation number
around the large central KAM island is estimated as
\begin{equation}
N_s = -log_3(\alpha_1) - log_3(\alpha_2) + 2
\end{equation}
This equation holds as long as at least one of the
 development parameters is sufficiently small. Compare with the ternary example
(1,1/3) in 
section 4.4: Count the number of cells around the central island (it is 3),
plug the two development parameters into Eq.6 and obtain again the number 3.

\section{Conclusions}
% Ab. 6

We give a systematic construction of an ( in general approximate ) partition
of an incomplete horseshoe describing the outer hyperbolic component of the
homoclinic tangle. This is exactly the part of the chaotic set 
which is most relevant for chaotic
scattering. Therefore it is essential that it leads to a branching tree 
coinciding with the one we extract from scattering functions. This is
guaranteed
by only using unstable manifolds of the outer fixed points and eventually
their limits as cell boundaries.

Of course, the ( iterated ) image or preimage of our partition would be
a completely equivalent partition. And the iterated preimages would also have
stable manifolds of the outer fixed points as division lines. This
demonstrates
that it is not so important to use unstable manifolds of the outer fixed
points
as division lines, it is important to use only any invariant manifolds of 
these particular points and not to use artificial division lines which do not
correspond to a change of the topology of the scattering trajectories.

The resulting symbolic dynamics indicates in which order scattering
trajectories
from the corresponding intervals of continuity of the scattering functions
visit
the various partition cells of the horseshoe. Therefore it is the appropriate
partition to connect scattering dynamics with the chaotic dynamics in the
interaction region, and it is the symbolic dynamics which we should
reconstruct
from scattering data when dealing with the inverse chaotic scattering problem.

{\bf Acknowledgement}: The author A. E. gratefully acknowledges
discussions with L.E. Reichl at the University of Texas at Austin.

\clearpage
\centerline{\bf Figure Captions}

\noindent{\bf Fig.1.} Binary horseshoe with $\alpha=1/2$ for $A=3.4$. The
thick broken lines in b) show schematically where the division lines between
the cells \%, \& and $+$ are in the limit to infinity.\\
\noindent{\bf Fig.2.} Binary horseshoe with $\alpha=1/2$ for $A=4$. In
Fig.(2b) the secondary tendril is the unstable manifold segment
between $A$ and $B$, indicated by the arrows.  \\
\noindent{\bf Fig.3.} Binary horseshoe with $\alpha=1$. \\
\noindent{\bf Fig.4.} Binary horseshoe with $\alpha=7/8$. \\
\noindent{\bf Fig.5.} Ternary asymmetric horseshoe with $\alpha's$ (1/3,1).
The thick broken lines in b) show schematically where the division lines
between the cells \%, \& and $+$ are in the limit to infinity.\\
\vspace{1cm}
\centerline{\bf Figures}

\vspace{1.45cm}
\begin{figure}[h]
\begin{centering}
\leavevmode
\epsfxsize=0.7\linewidth
\epsfbox{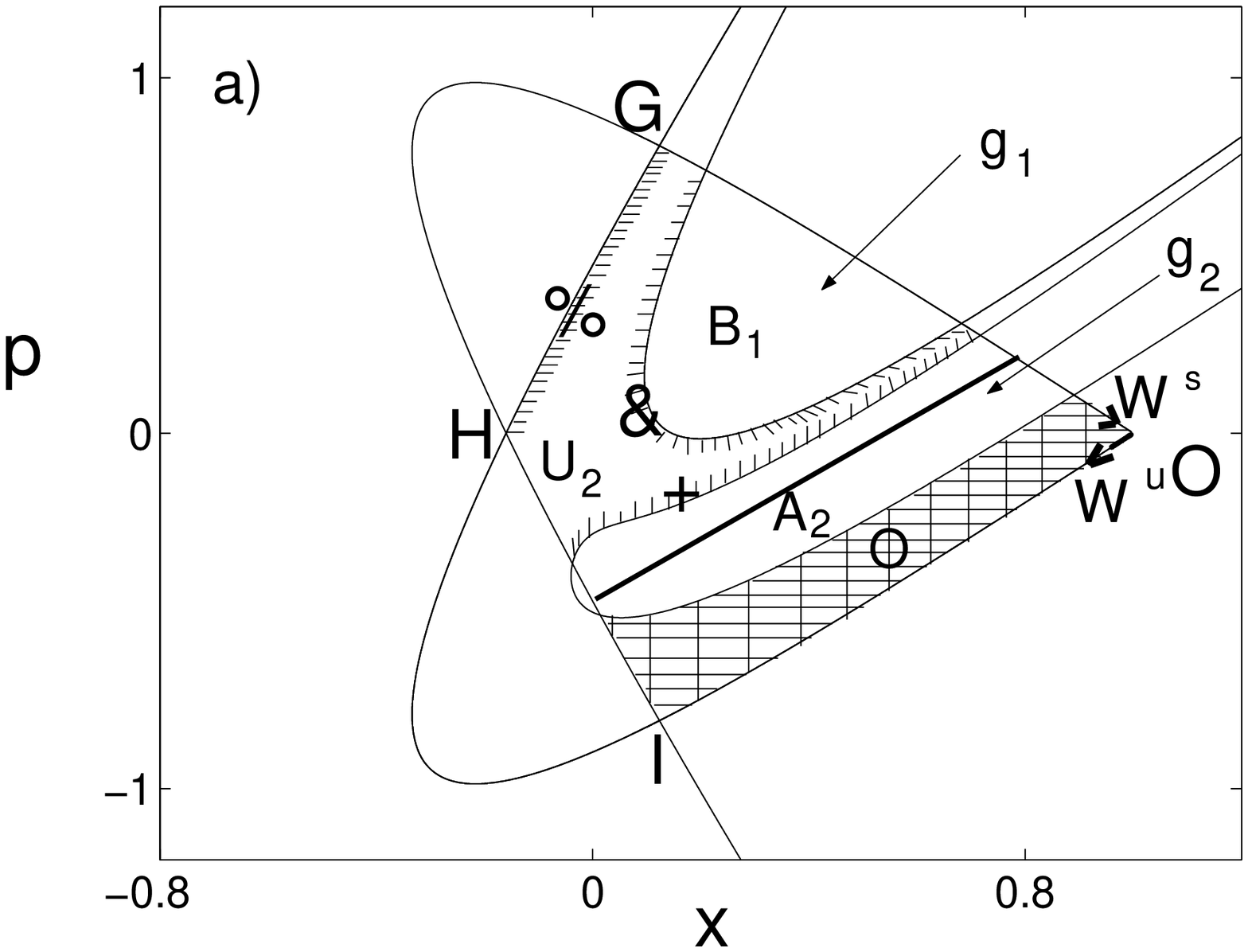}
\end{centering}
\begin{centering}
\epsfxsize=0.7\linewidth
\epsfbox{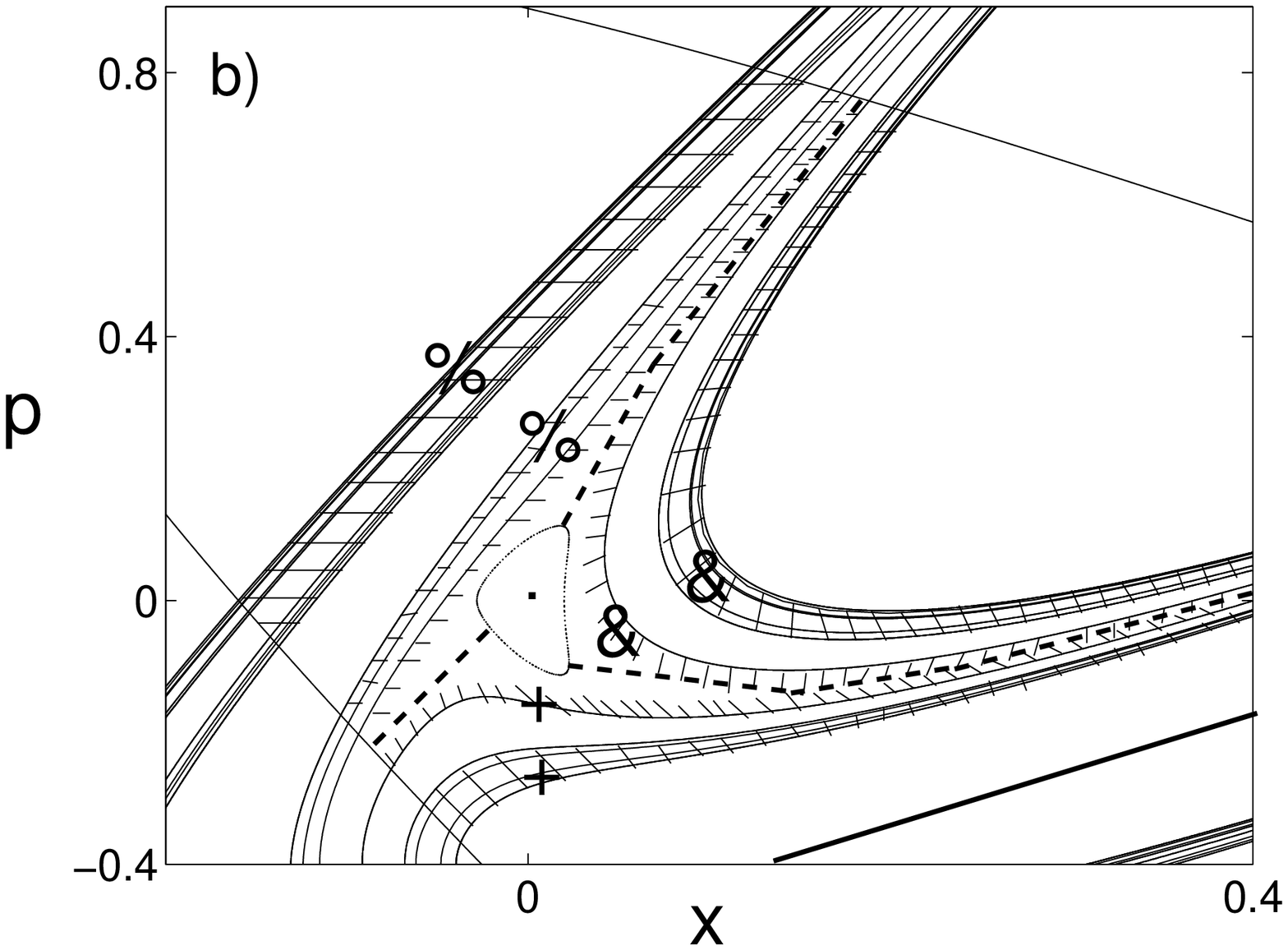}
\caption{}
\label{fig:binary}
\end{centering}
\end{figure}
\begin{figure}[h]
\begin{centering}
\leavevmode
\epsfxsize=0.7\linewidth
\epsfbox{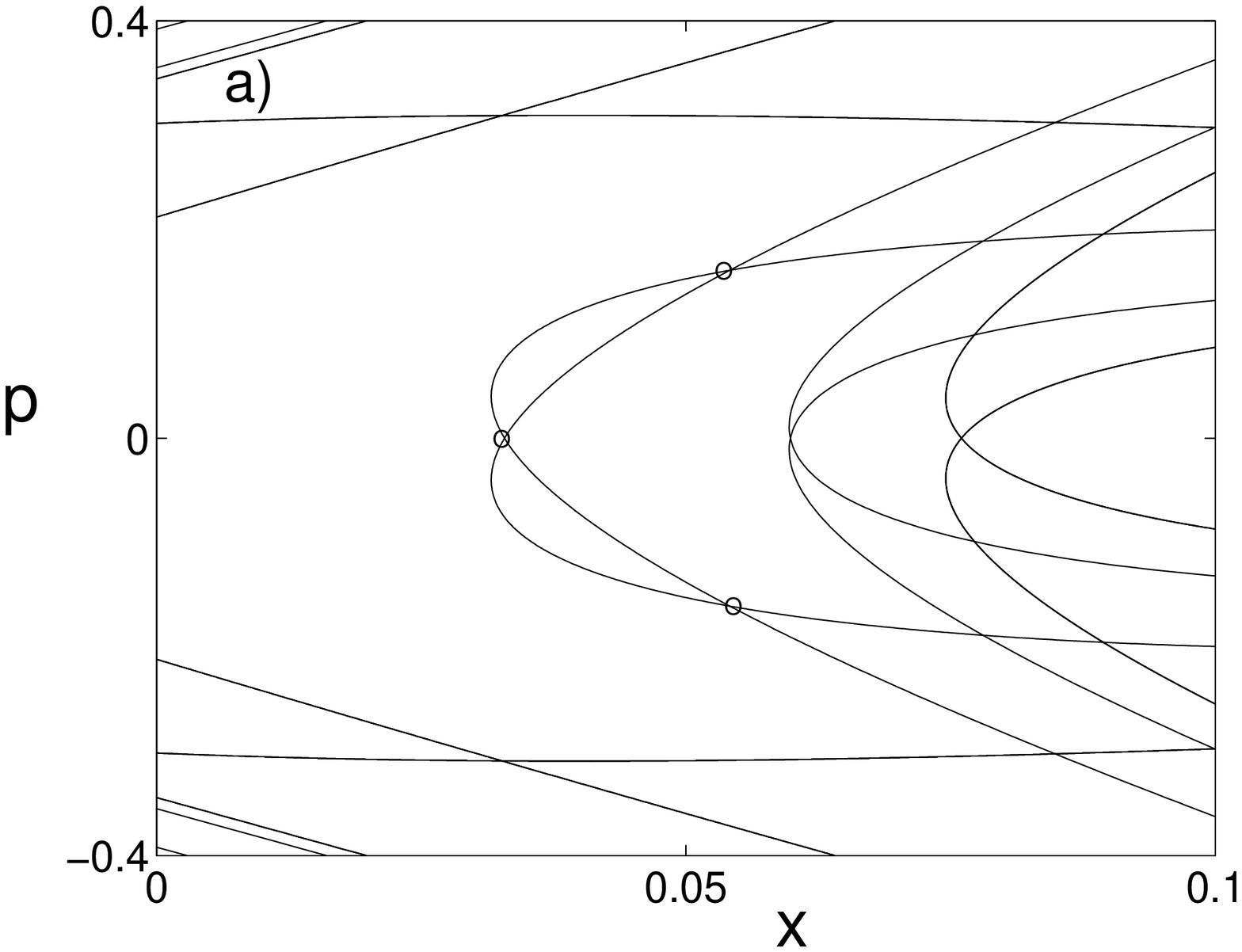}
\end{centering}
\begin{centering}
\epsfxsize=0.7\linewidth
\epsfbox{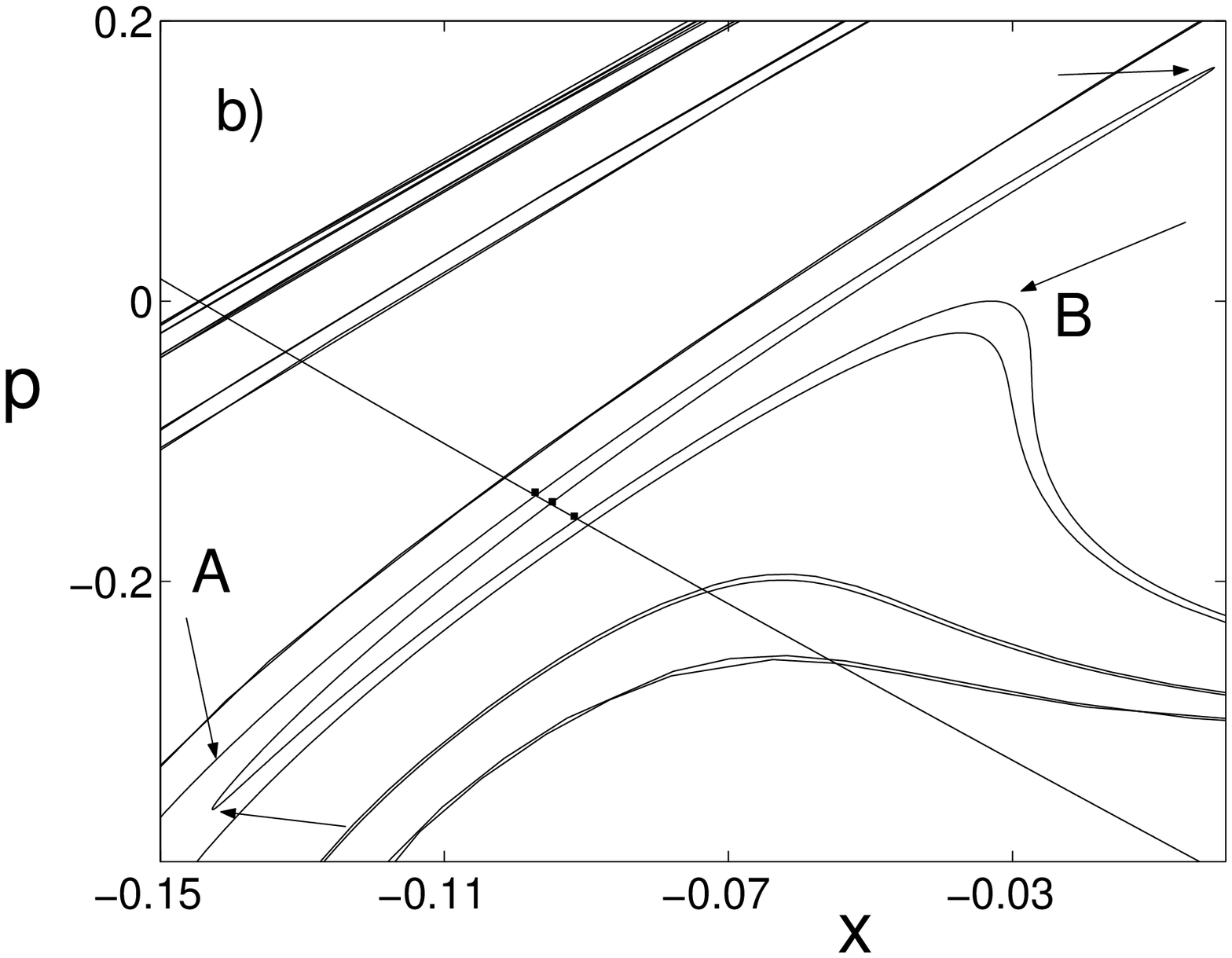}
\caption{}
\end{centering}
\label{fig:binary4}
\end{figure}
\begin{figure}[h]
\begin{centering}
\leavevmode
\epsfxsize=0.7\linewidth
\epsfbox{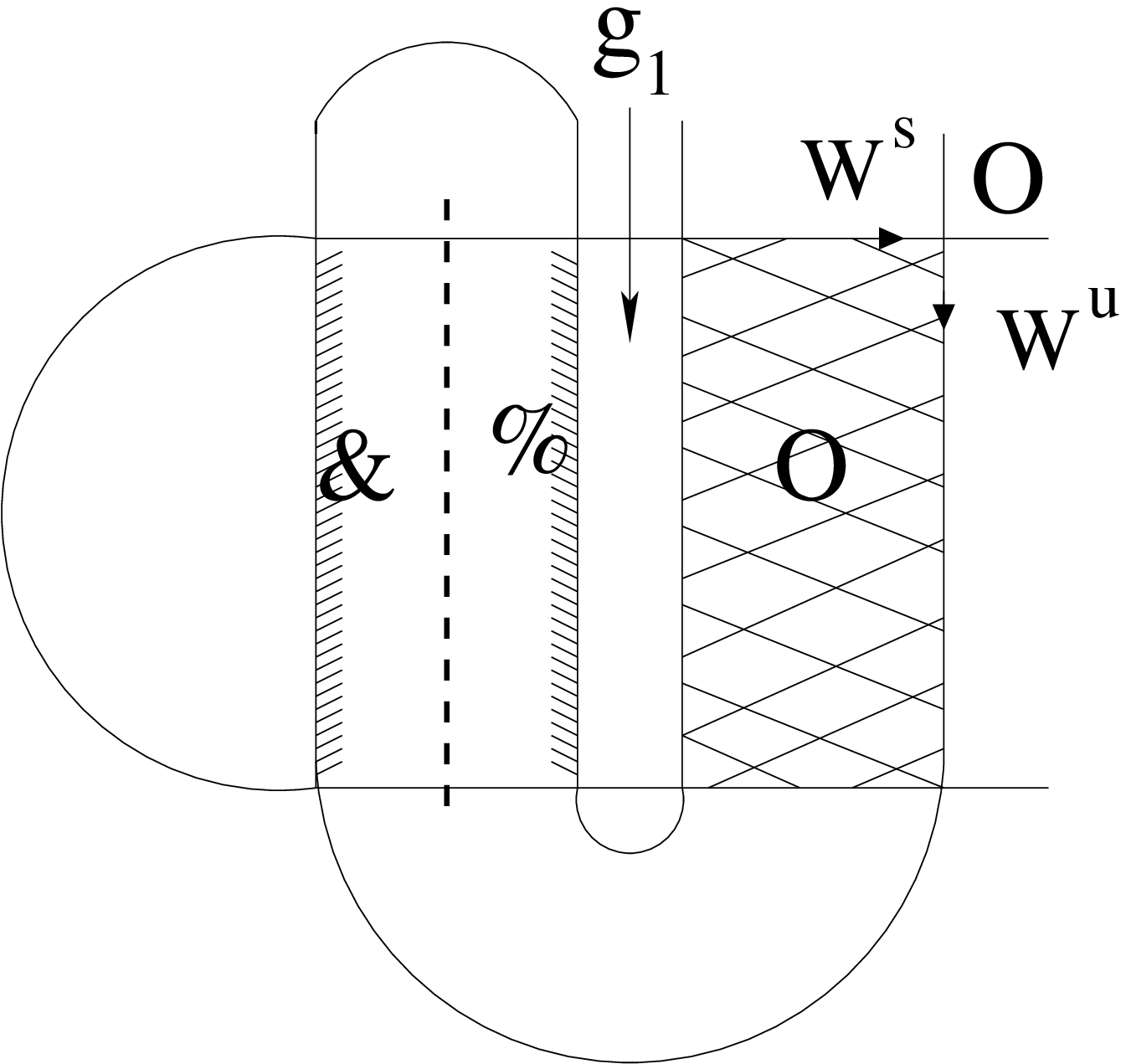}
\end{centering}
\caption{}
\label{fig:binary1}
\end{figure}
\begin{figure}[h]
\begin{centering}
\leavevmode
\epsfxsize=1\linewidth
\epsfbox{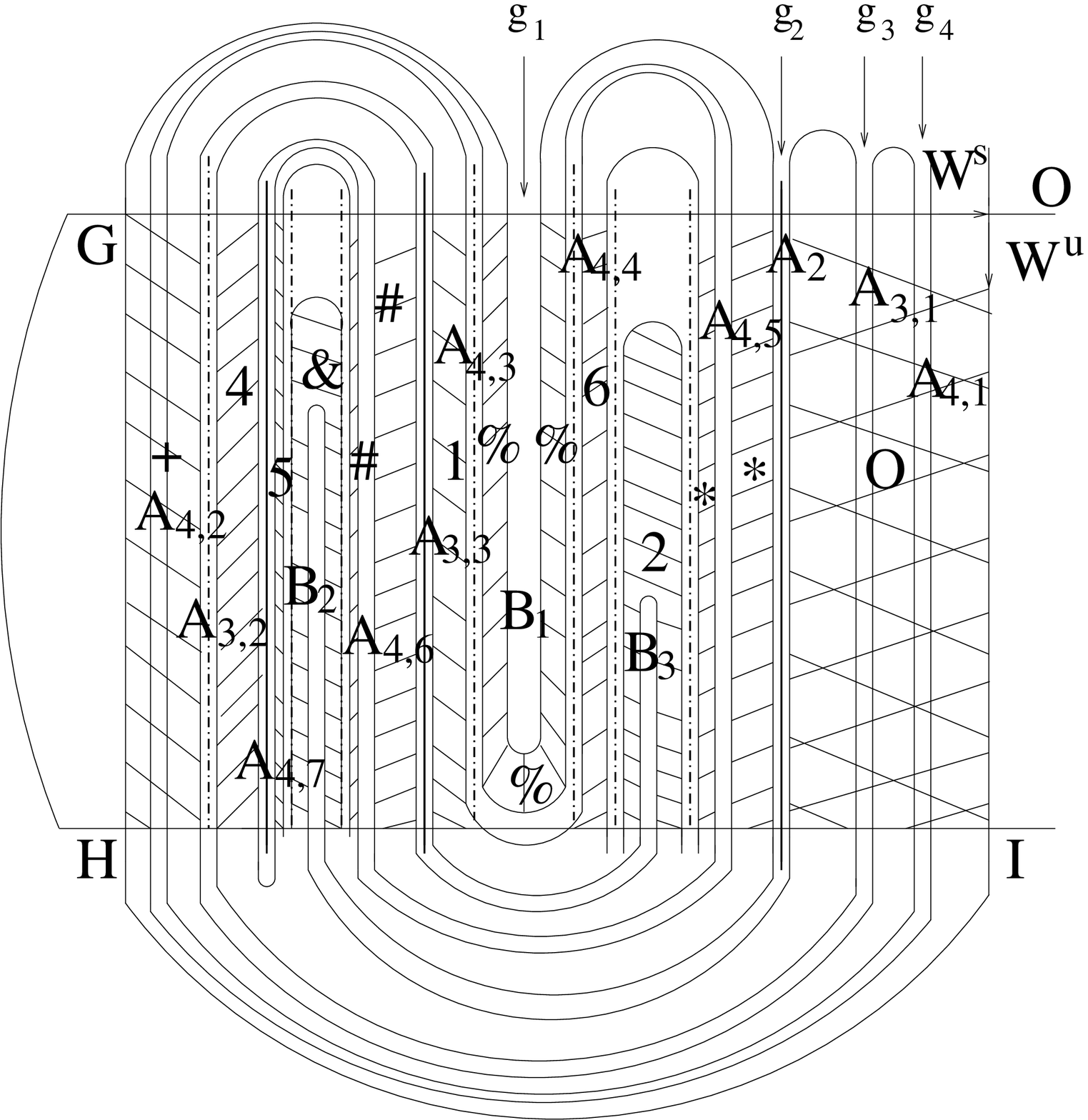}
\end{centering}
\caption{}
\label{fig:binary78}
\end{figure}
\begin{figure}[h]
\begin{centering}
\leavevmode
\epsfxsize=0.9\linewidth
\epsfbox{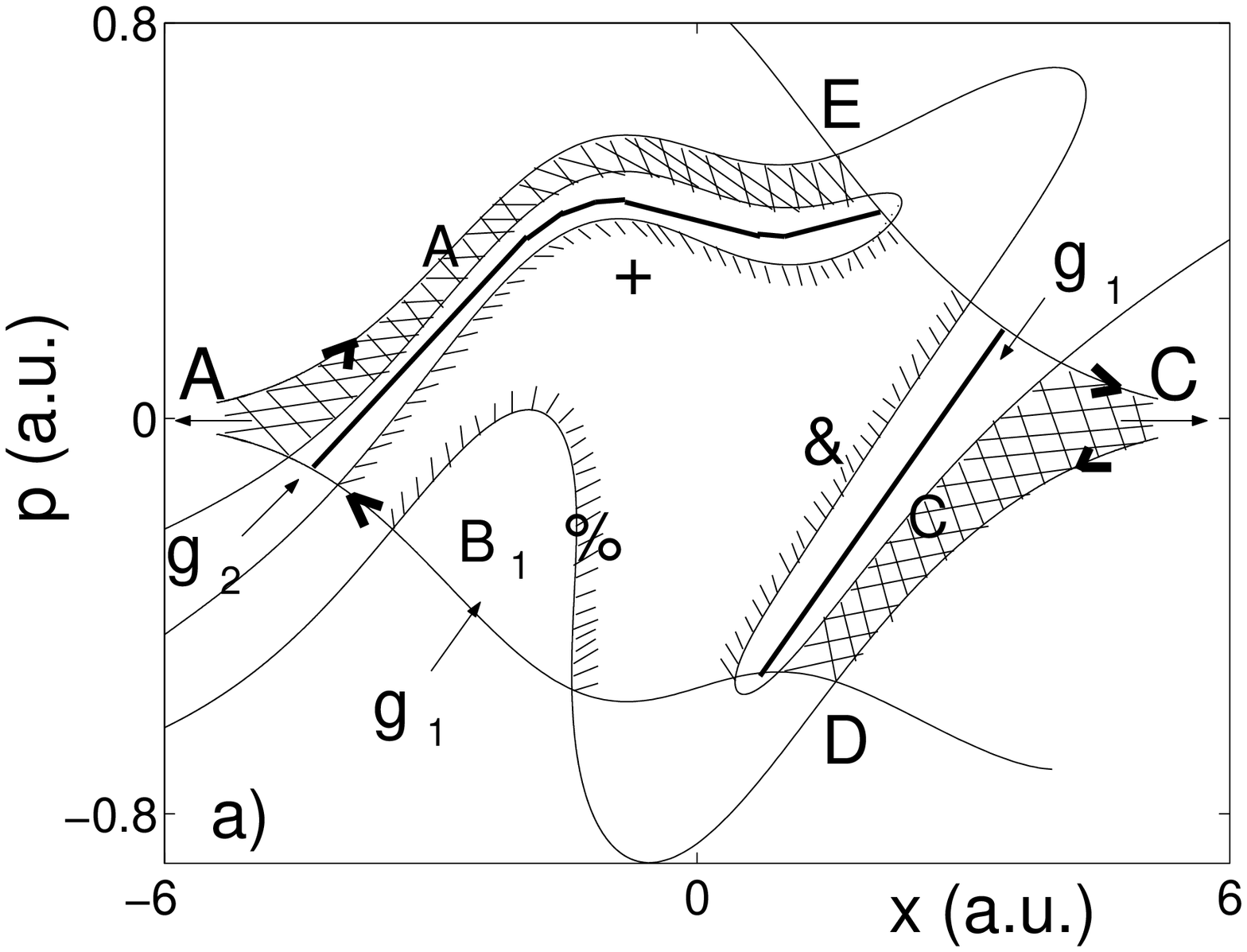}
\end{centering}
\begin{centering}
\epsfxsize=0.9\linewidth
\epsfbox{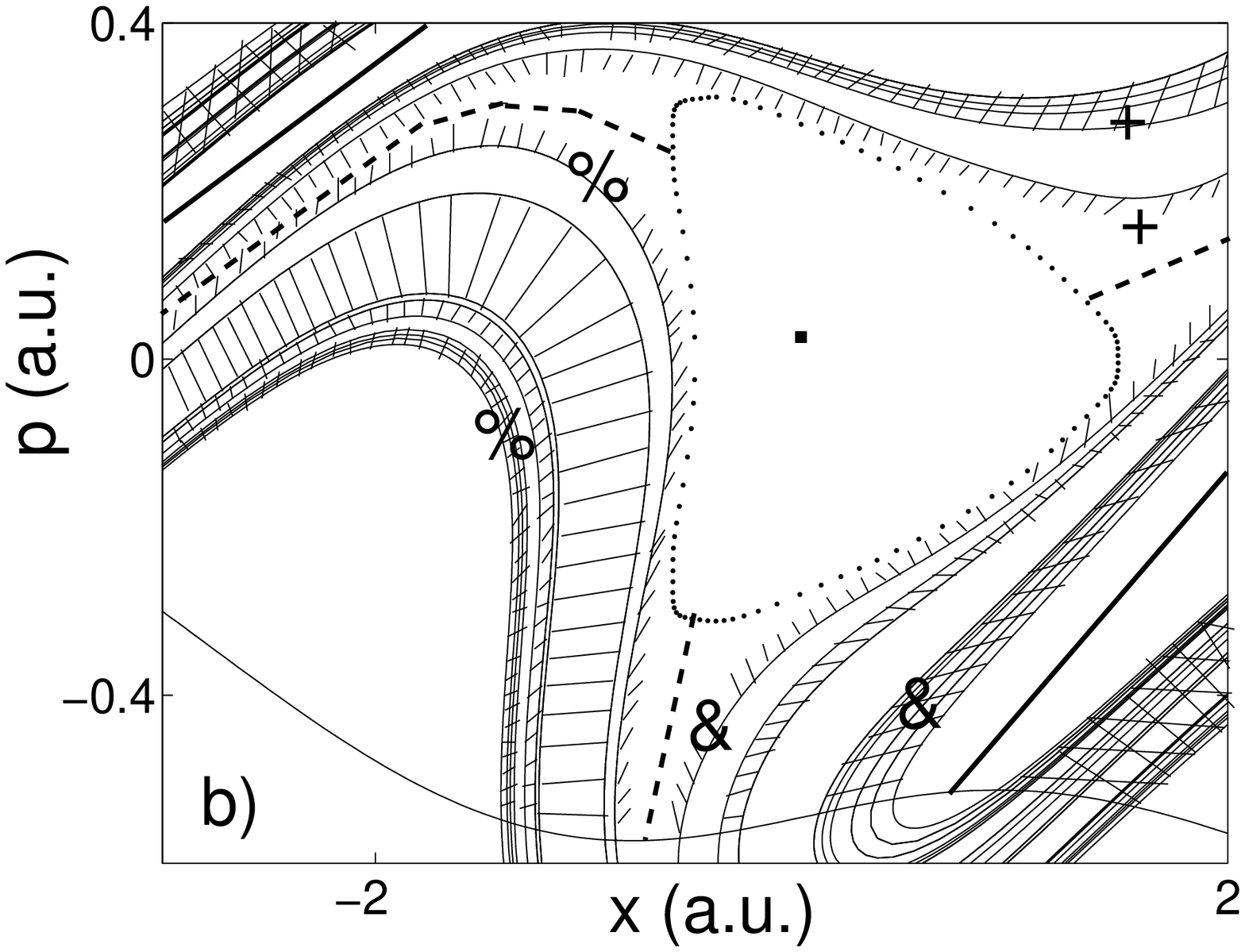}
\caption{ }
\label{fig:Fig4}
\end{centering}
\end{figure}

\end{document}